\documentstyle[12pt]{article}
\input{psfig}
\input{epsf}
\def\mc#1 {\multicolumn{1}{|c|}{#1}}
\begin{document}
\vspace*{2.0 cm}
\begin{center}
{\Large \bf Measurement of the Proton Structure Function ${F_2}$ \\
      at low ${x}$ and low ${Q^2}$ at HERA}\\
\vspace{2 cm}
{\large ZEUS Collaboration}
\end{center}
\vspace{5 cm}

\begin{abstract}
We report on a measurement of the proton structure function $F_2$ in the
range $3.5\times10^{-5}\leq x \leq 4\times10^{-3}$ and 1.5 ${\rm GeV^2} \leq
Q^2 \leq15$ ${\rm GeV^2}$ at the $ep$ collider HERA operating at a
centre-of-mass energy of $\sqrt{s} = 300$ ${\rm GeV}$.  The rise of $F_2$
with decreasing $x$ observed in the previous HERA measurements persists in
this lower $x$ and $Q^2$ range.  The $Q^2$ evolution of $F_2$, even at the
lowest $Q^2$ and $x$ measured, is consistent with perturbative QCD.
\end{abstract}

\vspace{-17cm}
\begin{flushleft}
\tt DESY 95-193 \\
October 1995 \\
\end{flushleft}

\setcounter{page}{0}
\thispagestyle{empty}
\newpage

%
\def\3{\ss}
\textwidth 15.5cm
\footnotesize
\renewcommand{\thepage}{\Roman{page}}
\begin{center}
\begin{large}
The ZEUS Collaboration
\end{large}
\end{center}
\noindent
M.~Derrick, D.~Krakauer, S.~Magill, D.~Mikunas, B.~Musgrave,
J.~Repond, R.~Stanek, R.L.~Talaga, H.~Zhang \\
{\it Argonne National Laboratory, Argonne, IL, USA}~$^{p}$\\[6pt]
G.~Bari, M.~Basile,
L.~Bellagamba, D.~Boscherini, A.~Bruni, G.~Bruni, P.~Bruni, G.~Cara
Romeo, G.~Castellini$^{1}$, M.~Chiarini,
L.~Cifarelli$^{2}$, F.~Cindolo, A.~Contin, M.~Corradi,
I.~Gialas$^{3}$,
P.~Giusti, G.~Iacobucci, G.~Laurenti, G.~Levi, A.~Margotti,
T.~Massam, R.~Nania, C.~Nemoz, F.~Palmonari, \\
A.~Polini, G.~Sartorelli, R.~Timellini, Y.~Zamora Garcia$^{4}$,
A.~Zichichi \\
{\it University and INFN Bologna, Bologna, Italy}~$^{f}$
% ...
\\[6pt]
\clearpage
\noindent
\hspace*{1mm}
$^{ 1}$ also at IROE Florence, Italy  \\
\hspace*{1mm}
$^{ 2}$ now at Univ. of Salerno and INFN Napoli, Italy  \\
\hspace*{1mm}
$^{ 3}$ supported by EU HCM contract ERB-CHRX-CT93-0376 \\
\hspace*{1mm}
$^{ 4}$ supported by Worldlab, Lausanne, Switzerland  \\
\hspace*{1mm}
% ...
$^{21}$ now with OPAL Collaboration, Faculty of Physics at Univ. of
        Freiburg \\
% ...

\noindent
\begin{tabular}{lp{15cm}}
$^{a}$ & supported by the Natural Sciences and Engineering Research
         Council of Canada (NSERC) \\
$^{b}$ & supported by the FCAR of Qu\'ebec, Canada\\
$^{c}$ & supported by the German Federal Ministry for Education and
         Science, Research and Technology (BMBF), under contract
         numbers 056BN19I, 056FR19P, 056HH19I, 056HH29I, 056SI79I\\
$^{d}$ & supported by the MINERVA Gesellschaft f\"ur Forschung GmbH,
         and by the Israel Academy of Science \\
$^{e}$ & supported by the German Israeli Foundation, and
         by the Israel Academy of Science \\
$^{f}$ & supported by the Italian National Institute for Nuclear Physics
         (INFN) \\
$^{g}$ & supported by the Japanese Ministry of Education, Science and
         Culture (the Monbusho)
         and its grants for Scientific Research\\
$^{h}$ & supported by the Korean Ministry of Education and Korea Science
         and Engineering Foundation \\
$^{i}$ & supported by the Netherlands Foundation for Research on Matter
         (FOM)\\
$^{j}$ & supported by the Polish State Committee for Scientific
         Research, grants No.~115/E-343/SPUB/P03/109/95, 2P03B 244
         08p02, p03, p04 and p05, and the Foundation for Polish-German
         Collaboration (proj. No. 506/92) \\
$^{k}$ & supported by the Polish State Committee for Scientific
         Research (grant No. 2 P03B 083 08) \\
$^{l}$ & partially supported by the German Federal Ministry for
         Education and Science, Research and Technology (BMBF) \\
$^{m}$ & supported by the German Federal Ministry for Education and
         Science, Research and Technology (BMBF), and the Fund of
         Fundamental Research of Russian Ministry of Science and
         Education and by INTAS-Grant No. 93-63 \\
$^{n}$ & supported by the Spanish Ministry of Education and Science
         through funds provided by CICYT \\
$^{o}$ & supported by the Particle Physics and Astronomy Research
         Council \\
$^{p}$ & supported by the US Department of Energy \\
$^{q}$ & supported by the US National Science Foundation
\end{tabular}

\newpage
\pagenumbering{arabic}
\setcounter{page}{1}
\normalsize

\vspace*{0.5 cm}
\section{Introduction}

The measurement of deep inelastic scattering (DIS), $ep \rightarrow eX$,
at HERA has shown a rapid rise of the proton structure function
$F_2(x,Q^2)$ with decreasing $x$ for
$x \le 10^{-2}$~\cite{b:ZEUS_F2,b:H1_F2}.
The corresponding increase of the virtual photon-proton cross section
$\sigma_{tot}^{\gamma^* p}$ with the centre-of-mass energy
$W$ is much stronger
than that of $\sigma_{tot}^{\gamma p}$
for real photons~\cite{b:sigtot_photoprod,b:H1_photoprod}.
The slower increase of the real photoproduction cross section is
consistent with the energy behaviour of
hadron-hadron total cross sections.
In perturbative QCD the rise of $F_2$ at low $x$ is ascribed to
an increase in the sea quark density~\cite{b:ZEUS_glu,b:H1_glu}, and the
significance thereof
is discussed extensively in the literature~\cite{b:DIS}.
One of the important questions is how far perturbative QCD
retains its validity as one probes large parton densities.
Until now the HERA measurements of $F_2$ have covered $Q^2$ values above
4.5 ${\rm GeV}^2$. It is of great interest to extend the $F_2$
measurement to lower $Q^2$ values and study the low $x$ behaviour
in the transition region between photoproduction and DIS.

In this paper, $F_2$ measurements,
from $e^+ p$ DIS data, at $x \ge 3.5 \times 10^{-5}$ and at
$Q^2$ values as low as 1.5~${\rm GeV}^2$
are reported. Access to such low $Q^2$
values is achieved in two different ways. For the first analysis (SVX)
HERA is operated with the interaction point shifted in the proton
direction in order to improve the acceptance for small positron
scattering angles. The second analysis (ISR) uses DIS events from the
nominal interaction point with initial state photon radiation
where the radiated photon is detected. These events effectively have a
lower initial positron beam energy and thus for a fixed acceptance of
the  positron
scattering angle events with smaller values of $Q^2$ can be reconstructed.
The reconstruction of events having positrons scattered
at small angles is improved compared to the 1993 analysis by the addition
of the small angle rear tracking detector (SRTD).

\section{Experimental Conditions}

\subsection{HERA Running Conditions}

The data were taken with the ZEUS detector at HERA in 1994.
HERA operated with 153 colliding bunches of 820~GeV protons
and 27.5 GeV positrons.
Additional unpaired positron and proton bunches circulated,
which are used to determine beam related background.
The root mean square of the proton bunch length was approximately 20 cm while
the positron bunch length was negligible in comparison, leading to
an interaction length having a root mean square of 10 cm.
For the SVX analysis, the mean interaction vertex was moved
from $Z$ = 0 to $Z$ = 67 cm~\footnote{The ZEUS coordinate system
is defined as right handed with the $Z$ axis pointing in the proton beam
direction, and the $X$ axis horizontal, pointing towards the centre of
HERA. The origin is at the nominal interaction point.}.
Approximately 5\% of the proton current was contained
in satellite bunches, which were shifted by 4.8~ns with respect
to the primary bunch crossing time, resulting in a fraction of
the $ep$ interactions occurring on average +72~cm upstream of the
primary position.

The SVX analysis is based on an integrated luminosity of 58~${\rm nb^{-1}}$
collected while HERA operated with the shifted interaction point.
The ISR analysis uses 2.5~${\rm pb^{-1}}$ of data
taken with HERA operating with the interaction point at
the nominal position, $Z = 0$.

\subsection{The ZEUS Detector}
A description of the ZEUS detector can be
found in~\cite{b:sigtot_photoprod,b:Detector}.
The primary components used in these analyses are the
uranium-scintillator calorimeter (CAL)~\cite{b:CAL} and the
tracking detectors.  The calorimeter covers
99.7\% of the total solid angle and is subdivided
into electromagnetic (EMC) and hadronic (HAC) sections with typical cell sizes
of $5 \times 20$ cm$^2$ ($10 \times 20$~cm$^2$
in the rear calorimeter (RCAL), i.e. in the positron beam direction)
and $20 \times 20$~cm$^2$ respectively.
The calorimeter has
an energy resolution of
$\sigma/E$~=~18\%/$\sqrt{E ({\rm GeV})}$
for electrons and
$\sigma/E$~=~35\%/$\sqrt {E({\rm GeV})}$ for hadrons,
as measured in test beams.
The timing resolution of a calorimeter cell is better than
$\sigma_t$~=~1.5/$\sqrt {E({\rm GeV})}$~$\oplus$~0.5~ns
($\oplus$ denotes addition in quadrature).

The tracking system consists of a vertex detector (VXD)~\cite{b:VXD}
and a central
tracking chamber (CTD)~\cite{b:CTD}  enclosed in a 1.43 T solenoidal
magnetic field. The interaction vertex
is measured with a resolution along (transverse to) the beam direction
of 0.4~(0.1)~cm.

The position of positrons scattered close to the positron beam direction
is determined by the SRTD which is attached to the front face of the RCAL.
The SRTD consists of two planes of
scintillator strips, 1 cm wide and 0.5 cm thick,
arranged in orthogonal directions and read out
via optical fibres and photomultiplier tubes.
It covers the region
of 68~$\times$~68~${\rm cm^2}$ in $X$
and $Y$ and is positioned at $Z = -148$~cm.
A hole of 20~$\times$~20~${\rm cm^2}$ at the centre of the RCAL and SRTD
accommodates the beampipe.
The SRTD is able to clearly resolve single minimum ionising particles (mip)
and has a position resolution of 0.3~cm.
The time resolution is better than 2 ns for a minimum ionising particle.

The luminosity is measured
via the positron-proton brems\-strahlung process, $ep \rightarrow e \gamma p$,
using a lead-scintillator calorimeter (LUMI)~\cite{b:LUMI}
which accepts photons at angles~$\le$~0.5~mrad with respect to the beam
axis.
The LUMI photon calorimeter is also used to measure the energy and position
of photons from initial state radiation in DIS events.
It is positioned at $Z = -107$~m and, under test beam conditions,
has an energy resolution of $\sigma/E$~=~18\%/$\sqrt{E~({\rm GeV})}$.
In its operating position it is shielded from synchrotron radiation
by a carbon-lead filter and has an energy resolution of
$\sigma/E$ = 26.5\%/$\sqrt {E ({\rm GeV})}$, as determined
from bremsstrahlung data. The position resolution is 0.2~cm.
In addition, an electromagnetic calorimeter positioned at
$Z = -35$~m is used for tagging scattered positrons at small angles.

\subsection{Triggering}

Events were filtered online by a three level trigger system~\cite{b:Detector}.
At the first level, DIS events were selected by requiring a logical AND
between two conditions based on energy deposits in the calorimeter.
The first condition was the presence
of an isolated electromagnetic energy deposit.
The EMC energy deposit was required to be greater than
2.5~GeV.  The corresponding HAC energy deposit was required to be either
less than
0.95~GeV or no more than a third of the EMC energy deposit.
The threshold values have been chosen
to give $>$99\% efficiency for positrons with energy greater than 5 GeV
as determined by Monte Carlo studies.
Further details of the algorithm can be found in~\cite{b:CALFLT}.
The second condition required that the EMC section have an
energy deposit greater
than 3.75 GeV. Background from protons interacting outside the detector
was rejected using the time measurement of the energy deposits from
downstream veto counters and the SRTD.

At the second level trigger (SLT),
background was further reduced using the measured
times of energy deposits and the summed energies from the calorimeter.
The events were accepted if
\begin{equation}
  \delta_{SLT} \equiv \sum_i E_i(1-\cos\theta_i) > 24
\:\:{\rm GeV} - 2E_{\gamma}
\end{equation}
where $E_i$ and $\theta_i$ are the energies and polar angles (with respect
to the primary vertex position) of calorimeter cells, and $E_{\gamma}$
is the energy deposit measured in the LUMI photon calorimeter.
For perfect detector resolution and acceptance,
$\delta_{SLT}$ is twice the positron beam energy (55~GeV)
for DIS events while for photoproduction events,
where the scattered positron escapes down the beampipe,
$\delta_{SLT}$ peaks at much lower values.

The full event information
was available at the third level trigger (TLT).
Tighter timing cuts as
well as algorithms to remove beam-halo muons and cosmic muons were
applied.
The quantity $\delta_{TLT}$ was determined in the same manner as for
$\delta_{SLT}$. The events were required to have
\begin{equation}
   \delta_{TLT} > 25 \:\:{\rm GeV} - 2E_{\gamma}.
\end{equation}
Finally, DIS events were accepted if
a scattered positron candidate of energy greater than 4~GeV was found.

\section{Monte Carlo Simulation}
\label{s:MC}

Monte Carlo (MC) event simulation is used to correct for detector
acceptance and smearing effects. The detector simulation is based on
the GEANT programme~\cite{b:GEANT} and incorporates our understanding
of the detector, the trigger and test beam results.
Neutral current DIS events are simulated to ${\cal O}(\alpha_s)$ using
the HERACLES programme~\cite{b:HCL} which includes first order
electroweak radiative corrections.
The hadronic final state is simulated using the
colour-dipole model including boson gluon fusion CDMBGF~\cite{b:CDM} as
implemented in ARIADNE~\cite{b:ARIADNE} for the QCD cascade and
JETSET~\cite{b:JETSET} for the hadronisation.
The ARIADNE model currently provides the best description of the observed
DIS nondiffractive hadronic final state~\cite{b:HAD_PAPERS}.
Diffractive events with a large rapidity gap as observed in the
data~\cite{b:LRG} are simulated within ARIADNE by assuming that the
struck quark
belongs to a colourless state having only a small fraction of the proton's
momentum. The parameters of the model are adjusted to be consistent with
recent ZEUS measurements~\cite{b:F2D}.
The MRSA~\cite{b:MRSA} parton density parameterisations, modified at
low $Q^2$ as described in~\cite{b:modifiedMRSA}, are used.
These parameterisations provide an adequate description of
previous ZEUS and H1 measurements~\cite{b:ZEUS_F2,b:H1_F2}.

The shape of the vertex distribution used in the simulation
is taken from nondiffractive photoproduction events. For such
events the vertex reconstruction efficiency is found to be
high and only weakly dependent on the $Z$ position of the interaction.

For the shifted vertex analysis, a sample of events corresponding to an
integrated luminosity of  $\sim 100$~${\rm nb^{-1}}$ was generated
with $Q^2 > 0.5$ ${\rm GeV}^2$.

The main source of background in the data sample for the SVX
analysis is due to photoproduction leading to the detection of a
fake scattered positron. Minimum bias photoproduction
events are simulated using PYTHIA~\cite{b:PYTHIA} with cross
sections according to the ALLM parameterisation~\cite{b:ALLM}.
Photoproduction events corresponding to an integrated luminosity of
110 ${\rm nb^{-1}}$ were generated with a photon-proton centre-of-mass
energy
$W\,\lower 2pt \hbox{$\stackrel{>}{\scriptstyle\sim}$ }\!190\,{\rm GeV}$.
Events with smaller $W$ values do not contribute to the photoproduction
background.

For the ISR analysis, a DIS sample corresponding to
an integrated luminosity of 2.7~${\rm pb^{-1}}$, with an initial
state photon energy above 3 GeV, was generated
with $Q^2 > 0.5$ ${\rm GeV}^2$.

\section{Kinematic Reconstruction}

\label{s:kinematic_reconstruction}
In deep inelastic scattering,
$e(k) + p(P) \rightarrow e (k^\prime) + X$,
the proton structure functions are expressed in terms
of the negative of the four-momentum transfer squared, $Q^2$,
and Bjorken $x$.
In the absence of QED radiation,
\begin{equation}
\label{e:q2}
Q^2 = - q^2 = -(k-k^\prime)^2,
\end{equation}
\begin{equation}
\label{e:x}
x = \frac{Q^2}{2P \cdot q},
\end{equation}
where $k$ and $P$  are the four-momenta of the
incoming particles and $k^\prime$ is the four-momentum of the scattered
lepton. The square of the centre-of-mass energy is denoted by $s$.
The fractional energy transferred to the proton
in its rest frame is $y = Q^2/(sx)$.

In the SVX analysis, $Q^2$ and $x$ are reconstructed from
the measured energy, $E_e^\prime$,  and
scattering angle, $\theta_e$ of the positron,
\begin{equation}
\label{e:q2el}
Q^2_{e} = 4E_eE_e^\prime \cos^2(\theta_e/2),
\end{equation}
\begin{equation}
\label{e:xel}
x_{e} = \frac{E_e E_e^\prime \cos^2(\theta_e/2)}
         {E_p (E_e - E_e^\prime \sin^2(\theta_e/2))},
\end{equation}
where $E_e$ and $E_p$ are the positron and proton beam energies
and the scattered positron angle is measured with respect to the
positive $Z$ direction.

The ISR sample is selected by requiring that a photon
with energy $E_\gamma$ be detected in the LUMI photon calorimeter.
The variables $Q^2$ and $x$ are determined
using equations~\ref{e:q2el} and~\ref{e:xel}
replacing $E_e$ with $E_e-E_\gamma$, treating the virtual positron
as a real positron, which is a good approximation
for the ISR analysis.

\subsection{Vertex Determination}
The vertex coordinates are determined from tracks reconstructed
with the CTD and VXD. The $Z$ coordinate is determined on an event-by-event
basis. Since the transverse sizes of the beams
are smaller than the resolutions for the $X$ and $Y$ coordinates of the
vertex, the beam positions averaged over the entire data sets
are used. For events which do not have a tracking vertex, the $Z$
coordinate is set to the primary position of the interaction point.
For events with a
tracking vertex, the resolution of $Z$ is $\pm 0.4$~cm over the entire
$Z$ range considered in these analyses.
At low $y$ the current jet is produced at small forward angles, resulting
in a reduced probability for vertex reconstruction.
The vertex distributions for the two analyses,
including events arising from the interaction of the
satellite bunch, are found to be well reproduced by the
MC simulation.

\subsection{Positron Identification and Efficiency}
The positron identification algorithm is based on a neural network
using information from the CAL
and is described elsewhere~\cite{b:NN}.
The efficiency of finding the scattered positron
is sensitive to details of the shower evolution, in particular to energy
loss in material between the interaction point and
the calorimeter. The efficiency was measured
using elastic QED Compton events $ep \rightarrow e\gamma p$
where the positron and photon are detected in the CAL
by exploiting the presence of two electromagnetic objects
and the overconstrained kinematics.
DIS data were also used to study the positron finding efficiency
by comparing different positron finding algorithms.
A correction based on the QED Compton study is used in the analyses.
The size of the correction to the efficiency obtained from the
standard ZEUS detector simulation programme
is 16\% at 8 GeV decreasing to 0\% at 18 GeV.
The efficiency of the identification algorithm when the scattered
positron has an energy of 10~GeV is 55\%, rising to 100\% above
energies of 18~GeV.  The uncertainty in the efficiency is accounted
for in the systematic errors (see Sect.~\ref{s:systematic_uncertainties}).

\subsection{Positron Position Measurement}

The impact point of the scattered positron at the calorimeter
is measured using the SRTD for the SVX data and part of the ISR data.
The position resolution of the SRTD is determined using positrons hitting
the calorimeter at the boundary of cells where the position
can be determined by the calorimeter with a resolution of less than
0.1~cm. The measured SRTD position resolution of 0.3~cm is well
reproduced by the MC simulation.

For events with a reconstructed vertex the scattering angle of
the positron, determined from the event vertex and the impact position
at the face of the SRTD, has a resolution of 1.7~mrad, while for events without
a tracking vertex the measured scattering angle has a
resolution of 4~mrad. For the ISR analysis, 7\% of the events are found
outside the SRTD.  For these events, the impact point is determined by
the RCAL resulting in an angular resolution of 7~mrad.

\subsection{Positron Energy Measurement}
In the analyses, $x$ and $Q^2$ are determined using the
corrected positron energy $E_e^\prime$ as described below.
For background rejection
and rejection of events with hard initial state radiation, we define,
in addition, the quantity $\delta$:
\begin{equation}
   \delta = \delta_h + \delta_e; \: \: \: \: \: \:
   \delta_h=\sum_h E_h(1-\cos\theta_h); \: \: \: \: \: \:
   \delta_e = E_e^\prime (1- \cos\theta_e),
\end{equation}
where $E_h$ is the energy deposited in the calorimeter cell $h$ and the
angle, $\theta_h$, is determined from the event vertex and the cell centre.
The sum excludes the calorimeter cells belonging to the scattered positron.
$\delta_e$ is calculated using the corrected positron energy.

In the $Q^2$ range of the present analyses the scattered
positron traverses typically two radiation lengths of passive
material before reaching the calorimeter,
thereby losing a significant amount of energy.
The correlation between the energy lost in the inactive material in front
of the calorimeter and the energy deposited in the
SRTD is used to correct the calorimeter energy measurement.

This correction can be determined from a data sample where the scattered
positron energy is known from kinematic constraints.
This is the case for DIS events with $x=E_e/E_p$, for which the scattered
positron energy
is equal to the incident positron beam energy. Events close to this
condition, called kinematic peak (KP) events, can be selected by the
requirement $\delta_h < 0.06\cdot E_e$.
The correction for lower energies can be obtained using
QED Compton events.
The energies of the positron and the photon can be predicted
precisely from the measurement of their scattering angles assuming the
transverse momentum of the scattered proton to be small.
Events from elastic DIS $\rho^{\rm o}$ production,
$ep \rightarrow ep\rho^{\rm o}$;~$\rho^{\rm o} \rightarrow \pi^+ \pi^-$
provide a way to check the SRTD energy correction. Here, the angle of
the positron is determined from the event vertex and the position
measurement at the face of the SRTD while the momenta of the
$\pi^+$ and $\pi^-$ are measured with the CTD.

\begin{figure}[p]
\begin{center}
\mbox{\psfig{file=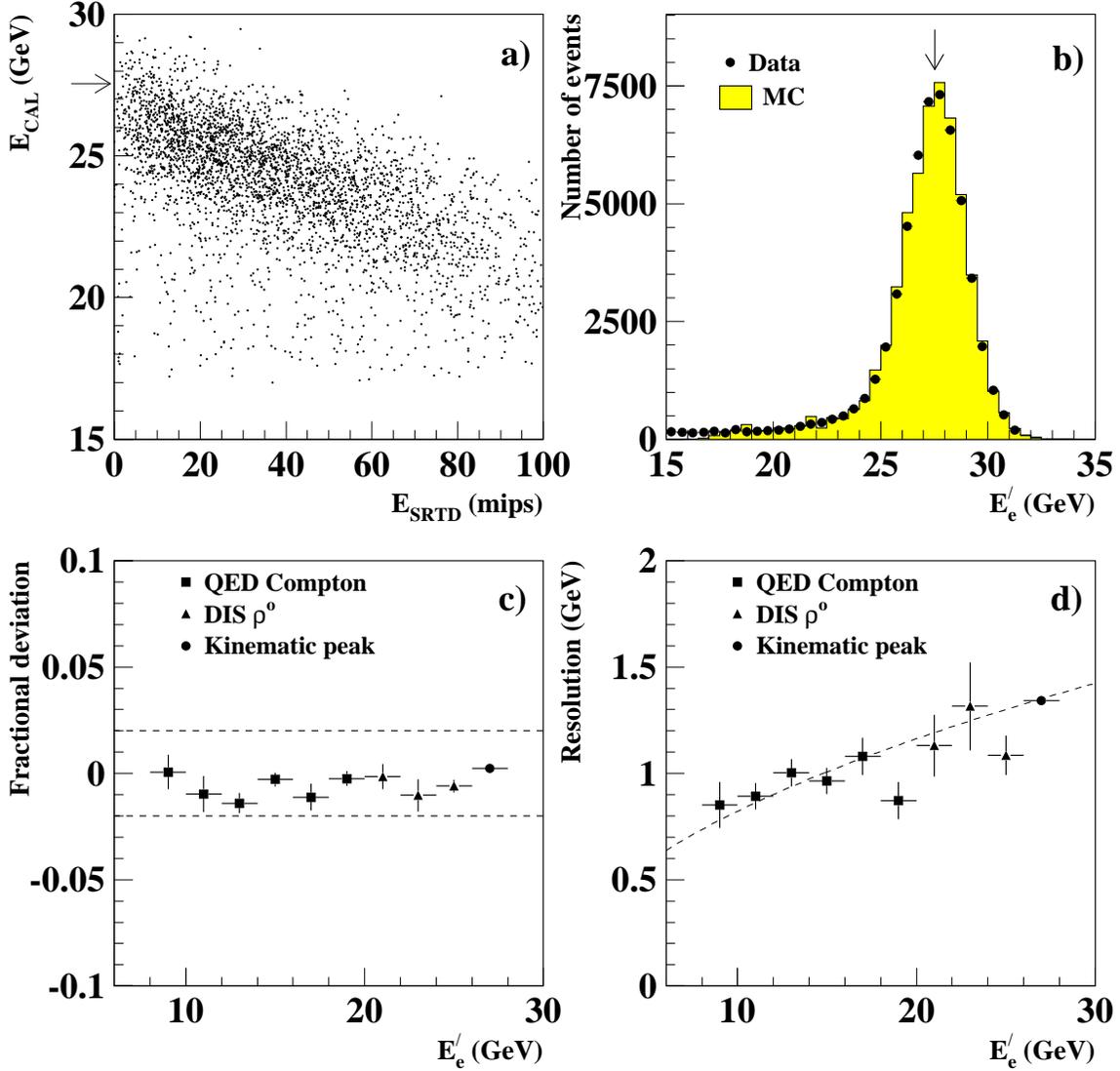,width=15cm}}
\end{center}
\caption[Energy correction.]
{\it
a) The correlation between the energy measured in the calorimeter and
the energy in the SRTD in units of mips (mean energy deposited by one minimum
ionising particle) for the kinematic peak (KP) events.
b) The distribution of the corrected positron energy for the data,
shown as the points, and the MC simulation, shown as the histogram,
for KP events. The arrows in a) and b) indicate the positron beam energy.
c) Measured fractional deviation between the mean corrected calorimeter
energy and the predicted energy as a function of the corrected energy.
d) The measured energy resolution of the calorimeter as a function of energy.
The curve corresponds to
26\%~$\cdot\sqrt{E({\rm GeV})}$, the resolution used in
the MC simulation. For c) and d), the data are results using QED Compton
(squares), DIS elastic $\rho^{o}$ (triangles) and KP (dots) events.
See text for details.
}
\label{f:ecorr}
\end{figure}

The SRTD energy correction is determined using the KP and QED Compton
events. A clear correlation between the energy measured in the calorimeter and
the energy deposited in the SRTD for KP events is observed (see
Fig.~\ref{f:ecorr}a).
Figure~\ref{f:ecorr}b compares the corrected positron energies of the
data with the MC simulation for KP events,
where it can be seen that the peak and width
of the data are well reproduced by the MC simulation.
The deviation of the mean corrected energy from the prediction of
the KP and QED Compton events is less than 2\% as shown in
Fig.~\ref{f:ecorr}c for positron energies between 9 and 27.5 GeV.
The point at the highest energy ($E_e^\prime$~=~27.5~GeV)
is obtained from the KP events while the lower energy points are determined
from the elastic QED Compton events.
For the DIS $\rho^{\rm o}$ data, the deviation of the mean corrected positron
energy
from the value expected from the $\rho^{\rm o} \rightarrow \pi^+ \pi^-$
measurement is less than 1\% (see Fig.~\ref{f:ecorr}c);
these data have not been used to determine the
SRTD energy correction. The resolution of the corrected positron
energy is shown in Fig.~\ref{f:ecorr}d and can be described by
$\sigma/E$~=~26\%/$\sqrt {E ({\rm GeV})}$.

\section{Analysis of the Shifted Vertex Data}

In the SVX analysis, $F_2$ is measured using data from
an integrated luminosity of
58~${\rm nb^{-1}}$
collected with the vertex shifted by +67~cm in $Z$.  The vertex
shift extends the acceptance of positrons to smaller scattering
angles and hence events with lower $Q^2$ can be reconstructed.

\subsection{Event Selection}
\label{s:event_selection}
The following cuts are used to select the DIS events for the SVX analysis.
\begin{itemize}
\item The scattered positron energy as obtained from the calorimeter
      and corrected by the energy measured in the SRTD is required to be
      greater than 10 GeV.
      This ensures a high efficiency of finding
      the scattered positron and removes high $y$ events which suffer from
      large photoproduction background.
\item The impact position of the scattered positron is
      required to be at least
      3~cm from the inner edge of the rear calorimeter.
      This cut ensures that the electromagnetic shower
      of the positron is well contained in the calorimeter.
      The impact position of the scattered positron is required to be
      within the acceptance of the SRTD and at least
      1 cm away from the edges of the SRTD.
\item The value of $\delta$ for the event is required to
      be within $35 < \delta < 65$~GeV in order to
      reduce photoproduction and beam-gas related background.
      This requirement also removes events with hard initial
      state radiation, thus reducing the radiative corrections.
      The hadronic energy measurement affects the event selection via
      this cut when $\delta_e < 35$~${\rm GeV}$. The simulation
      of $\delta_h$ in the MC is sensitive
      to the hadronisation scheme and the details of the hadronic energy loss
      in the inactive material.
      The mean $\delta_h$ was compared
      to $\delta_e$ as a function of $\delta_e$ and the MC
      simulation was found to reproduce the data adequately.
\item For events with a tracking vertex, the $Z$ coordinate of the vertex
      is required to be within $25\:{\rm cm}<Z<200\:{\rm cm}$,
      the acceptance being extended to larger $Z$ values to accommodate the
      events from the satellite bunch.
      If no vertex was reconstructed, the $Z$ coordinate is set to the
      primary interaction point.
\end{itemize}
A total of 13210 events pass the selection cuts.

\begin{figure}[p]
\begin{center}
\mbox{\psfig{file=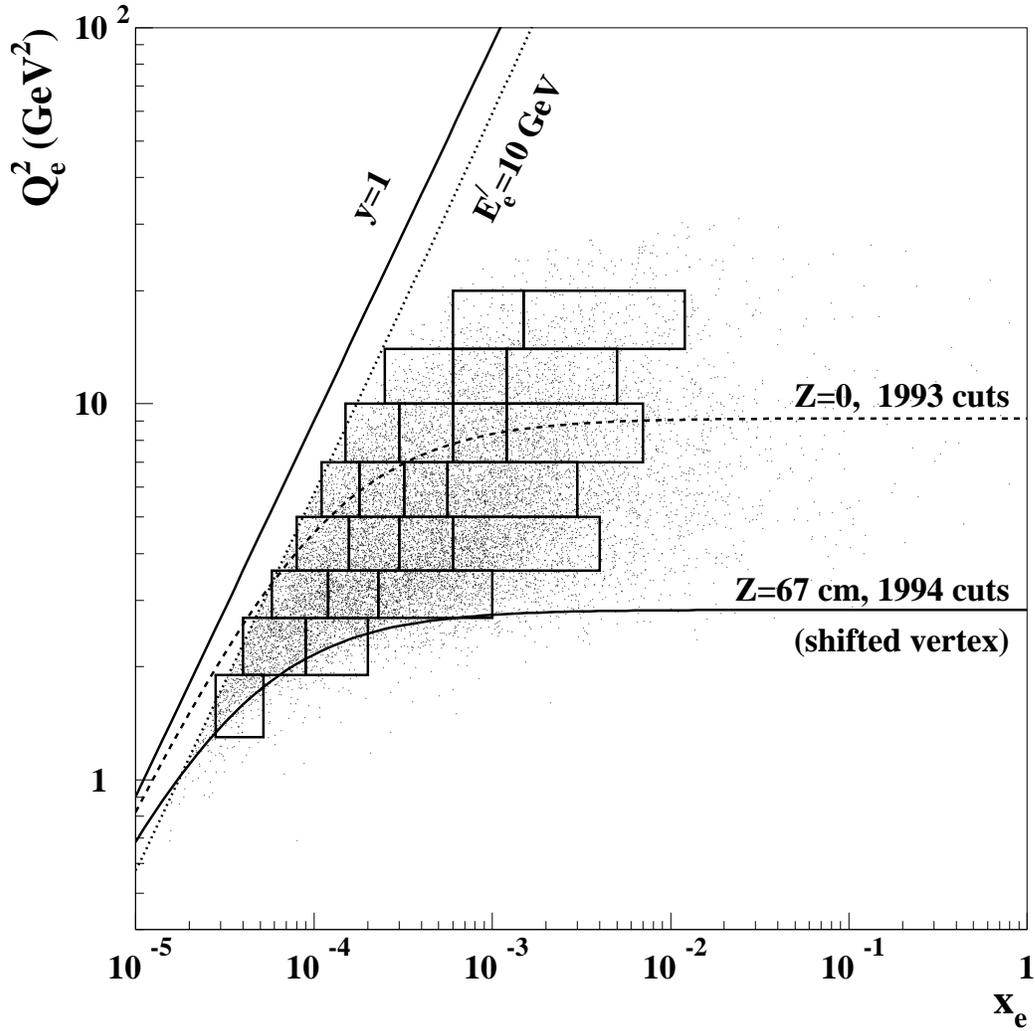,width=15cm}}
\end{center}
\caption[The extended region.]
{\it
The $x$-$Q^2$ distribution for events passing the selection
criteria from the 1994 SVX analysis.
The extension in the accepted region compared to the 1993 analyses
is shown between the dashed line labeled
``Z=0, 1993 cuts'' and the solid line labeled
``Z=67 cm, 1994 cuts (shifted vertex)''. See text for more details.
}
\label{f:fig_shifted_q2_x_extension}
\end{figure}

The distributions of the reconstructed $x$ and $Q^2$
for the selected events are
shown in Fig.~\ref{f:fig_shifted_q2_x_extension}.
The dashed line (Z=0, 1993 cuts) gives
the limit of the acceptance for the scattered positron corresponding to
the 1993 $F_2$ measurement using the ZEUS detector.
The acceptance limit in this analysis
is shown as the solid line (Z=67 cm, 1994 cuts).
The region between the dashed and solid lines shows the increased
acceptance obtained for the SVX data and by using the SRTD.

\subsection{Resolution of the Kinematic Variables and Bin Selection}
\label{s:resolution}
The selected events are binned as shown in
Fig.~\ref{f:fig_shifted_q2_x_extension}.
The sizes of the bins are determined by the resolution in $x$ and $Q^2$.
Within the selected bins, the resolution in $Q^2$
is found to be 9-12\% and the systematic shifts are less than 5\%
as determined from MC simulation.
At the lowest values of $x$, the resolution in $x$ is 20\% increasing to
85\% for larger $x$ values where larger bins are chosen to ensure a high
purity (defined as the number of events measured in the bin
which originated from the bin divided by the number of events
measured in the bin).
The average reconstructed values of $x$ are typically shifted by 6\%, with
shifts of up to 28\% occurring at the lowest $y$ values.
The bias in the bins results from the steepness of the distribution and
the radiative events.

The purity of the events in the bins is typically about 45\% and
is greater than 30\% for all bins.
The acceptance (defined as the number of measured events
originating from the bin divided by the number of events
generated in the bin) is typically 70\% except for the lowest
$Q^2$ bin where it is about 22\%, and for the bin with
1.9 ${\rm GeV^2} < Q^2 < 2.7$ ${\rm GeV^2}$ and
0.9 $\times 10^{-4} < x < 2.0$ $\times 10^{-4}$ where it is 27\%.
In these bins the purity is 50\% and 36\% respectively.

\subsection{Background Estimation}
The background from beam-gas related interactions
is estimated from events due
to the unpaired bunches of positrons and protons.
Events triggered from empty bunch crossings
are used to estimate the background from cosmic ray showers.
The surviving background events are subtracted statistically.
The $(x, Q^2)$ bin with the highest background has 2\% background.
In the other bins no events survive and the background is estimated to
be less than 1\%.

The largest contributor to background is
photoproduction, where an energy deposit in the calorimeter is
misidentified as a scattered positron.
This background is estimated in two ways.  The first method
uses the photoproduction MC simulation. The same selection procedure
as for DIS events is applied and the accepted photoproduction events are
statistically subtracted for each $(x, Q^2)$ bin.
The second method uses the $\delta$ distribution  to
estimate the background. The two estimates are found to
agree within statistical errors.
The photoproduction background is found to be significant only
in bins corresponding to the highest $y$ values where it amounts to 2\%.

\subsection{Determination of $F_2$}
\label{s:determine_f2}

In the $Q^2$ range of the present analysis, the effect from $Z^o$ exchange
is negligible and
the double differential cross section for single
virtual photon exchange in DIS
is given by
\begin{equation}
\label{e:f2}
\frac{d^2\sigma}{dx dQ^2} = \frac{2 \pi \alpha^2}{x Q^4}
              \left[ 2(1-y) + \frac{y^2}{1 + R} \right] F_2(x, Q^2)
              \left[1 + \delta_r(x,Q^2) \right],
\end{equation}
where $R$ is related to the longitudinal structure function, $F_L$, by
$R = F_L / (F_2 - F_L)$.
The correction to the Born cross section from radiative effects
is given by $\delta_r$. In the analysis, hard radiation collinear
with the final state positron is not resolved.  Furthermore,
the event selection criteria (see above) remove events with
hard radiation collinear with the beam positron. Thus,
in the kinematic range covered in this analysis, the
effective radiative correction is typically 10\% or less and is,
to a sufficient approximation, independent of the structure
function $F_2$.

An iterative procedure is used to extract the structure function
$F_2$. In the first step a bin-by-bin correction obtained from
the MC simulation using the parton distributions
given by MRSA\cite{b:MRSA} is applied to the data.
The result for $F_2$ from this first iteration is used for
a QCD fit using DGLAP equations~\cite{b:GLAP}
in next-to-leading order (QCD NLO fit)
very similar to that described
in~\cite{b:ZEUS_glu}.
The evolution uses massless quarks of three flavours in the proton,
and the charm quark coefficient functions from~\cite{b:REYA}
to ensure a smooth crossing of the charm threshold;
the charm contribution to $F_2$ is calculated in leading order only.
The NMC~\cite{b:NMC} data for $Q^2 > 4~{\rm GeV^2}$ are used to
constrain the fit at high $x$. The value of $R$ was taken using the QCD
prescription~\cite{b:AM} and
the parton distributions from the
QCD NLO fit. The effect of the $R$ correction on the $F_2$ values
is significant only in the highest $y$ bins where it is as much as 7\%.

The resulting QCD NLO fit parameters are used to reweight the MC
events, and the procedure is repeated, leading to a new estimate of
$F_2$.
The procedure is repeated until the $F_2$ values from two
consecutive iterations change by less than 0.5\%. The final result
is reached in three iterations.
It should be noted that the QCD NLO fit is used here only as a
parameterisation to obtain a
stable acceptance correction and not to perform a QCD analysis.

The statistical errors of the $F_2$ values are calculated from the
number of events measured in the bins and the statistical error on the
acceptance calculation from the MC simulation.
Since we are using bin-by-bin corrections, the correlations of statistical
errors between the $F_2$ measurements enter only via the finite statistics
of the MC sample.  The correlations are small given the relatively large
MC sample used in this analysis.
A correlation between the $F_2$ values of neighbouring bins
is present due to the acceptance and smearing corrections.
The sensitivity of the measured $F_2$ to this effect has been checked by using
the $F_2$ obtained from the
MRSA~\cite{b:MRSA} parton density parameterisations, modified at
low $Q^2$ as described in~\cite{b:modifiedMRSA},
for the acceptance
correction calculations.  The change
in the measured $F_2$ values is found to be small compared to
the statistical errors.

\begin{figure}[p]
\begin{center}
\mbox{\psfig{file=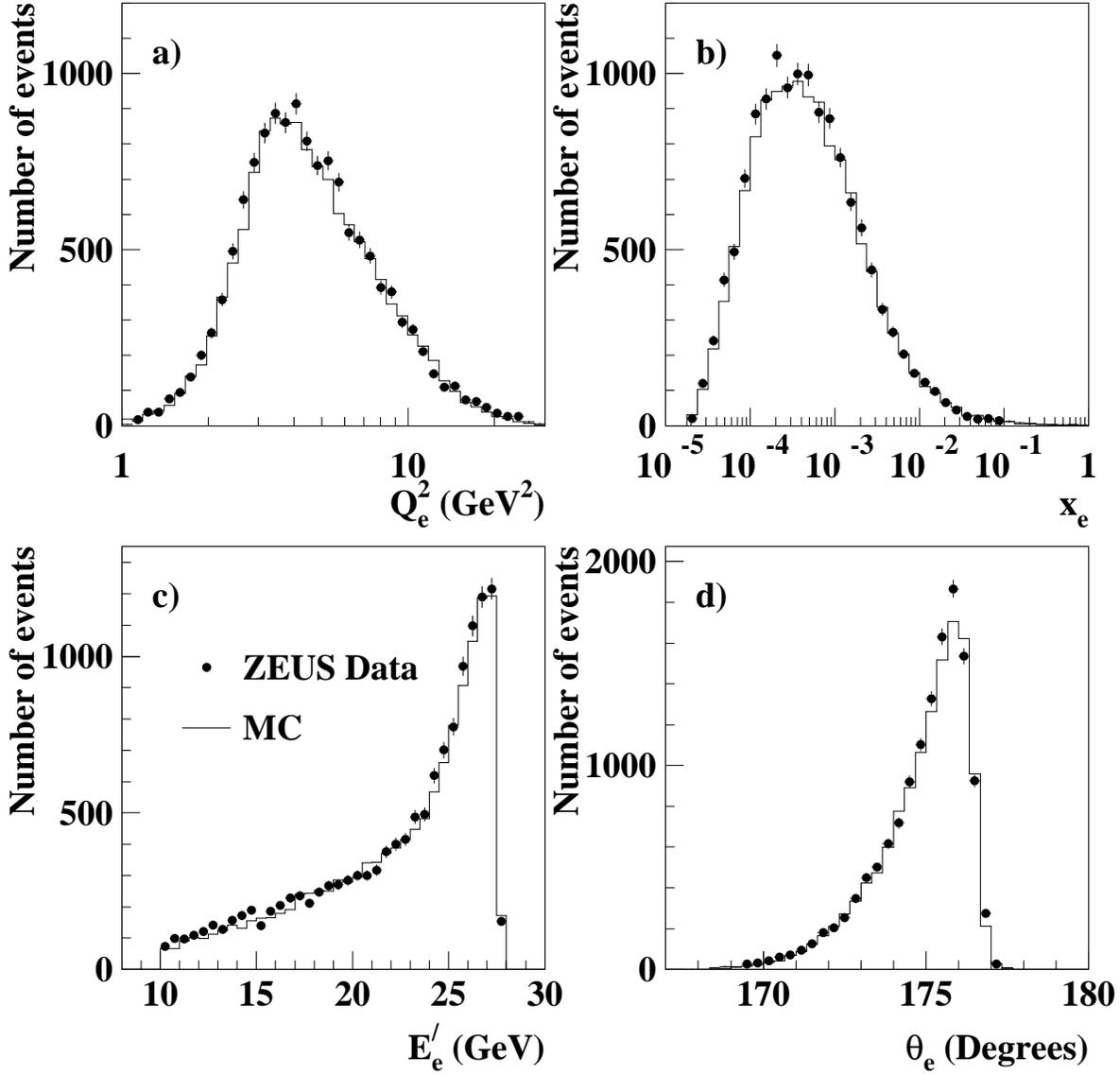,width=17cm}}
\end{center}
\caption[kkk]
{\it
a) The reconstructed $Q_e^2$ of the SVX event sample.
b) The reconstructed $x_e$ distribution of the SVX event sample.
c) The spectrum of the scattered positron energy.
d) The distribution of the positron scattering angles.
In the figures the data (dots) are compared with the
MC simulation (histograms).
All events with a reconstructed $Q^2_e > 1$~${\rm GeV^2}$
which pass the selection criteria described in
Sect.~\ref{s:event_selection} are shown.
The background has not been subtracted from the data.
The MC distributions have been reweighted using
the final $F_2$ parameterisation from the QCD NLO fit to the ZEUS data and
normalised to the luminosity of the data.
}
\label{fig_measured}
\end{figure}

The measured distributions of $Q^2$ and $x$ are shown in
Figs.~\ref{fig_measured}a and~\ref{fig_measured}b respectively.
The data are shown as the solid circles and the MC
simulation results, normalised to the
luminosity of the data and reweighted by the QCD NLO fit, are  shown as
histograms. The measured distributions of
the positron energy and angle are shown in
Figs.~\ref{fig_measured}c and~\ref{fig_measured}d.
All events which pass the selection criteria described in
Sect.~\ref{s:event_selection} with a reconstructed
$Q^2~>~1$~${\rm GeV}^2$ are shown, including those events
which fall outside the bins used for the analysis.
These distributions have not been corrected for background.
There is adequate agreement between data and simulation for both the
shape and normalisation of the distributions. The number of events,
and values of $R$ and $F_2$ at
$x$ and $Q^2$ values specified, which are chosen to be convenient
for comparisons with other $F_2$ measurements,
are given in Table~\ref{F2_table}.
The functional
form of the QCD NLO fit was used to calculate $F_2$ at these $x$
and $Q^2$ values.
Using alternate parameterisations for this correction
has a negligible effect on the quoted values of $F_2$.

\subsection{Systematic Uncertainties}
\label{s:systematic_uncertainties}
The systematic uncertainty of the $F_2$ values is determined by changing the
selection cut or analysis procedure and taking the
difference between the measured value of $F_2$ and the new value. Positive
and negative differences are separately added in quadrature to
obtain the total systematic error. For each change, the photoproduction
background estimated from the MC simulation is first subtracted.

The systematic uncertainties are grouped in the following categories:

\vspace{5mm}
\noindent
{\em Positron energy and finding efficiency}
\begin{itemize}
\item The parameters of the neural network positron finder are varied
      resulting in changes of $F_2$ by typically less than 4\%.
      In addition, an alternate positron finder, which has
      been used in previous analyses~\cite{b:ZEUS_F2}, is used as a
      check of the neural network finder. Consistent results are
      obtained in the region where the efficiencies of both finders are
      reasonably high.
\item In the MC simulation the reconstructed positron energy
      is increased and
      decreased by a linear function (2\% at 5 GeV and 1\% at 27.5 GeV).
      The magnitude of the shift represents our present understanding of the
      energy scale.
      These shifts have a 5\% effect on $F_2$ for the low $x$ bins and 10\%
      for the highest $x$ bins.
\item The positron identification efficiency is varied within the
      errors of its determination based on the QED Compton study. The effect
      on $F_2$
      is negligible except in the lowest $x$ bins where it is as much as 3\%.
      This also checks the effects of a possible energy dependence of the
      trigger efficiency.
\end{itemize}

\vspace{5mm}
\noindent
{\em Positron scattering angle}
\begin{itemize}
\item Changing the fiducial cut of the positron position
      at the inner edge of the
      calorimeter from 3 cm
      to 4 cm has a 3\% effect on $F_2$ for the lowest $x$ and $Q^2$ bins.
\item The reconstructed $Z$ vertex position
      in the MC simulation is shifted by
      $\pm$0.4~cm in $Z$.
      The effect on $F_2$ is less than 1.5\% for all the bins.
\item The positron position reconstructed by the SRTD is shifted by
      $\pm$0.2~cm in $X$ and $\pm$0.15~cm in $Y$.
      The size of the shifts are estimated from the uncertainty
      of the position of the SRTD relative to the beam position in $X$ and $Y$.
      These shifts have a 7\% effect on $F_2$ for the lowest $x$ bins and
      less than
      4\% for the other bins.
\end{itemize}

\vspace{5mm}
\noindent
{\em Additional systematic uncertainties}
\begin{itemize}
\item The photoproduction background estimate is changed by +100\%
      and $-50$\% resulting in a
      2\% effect on $F_2$ for the lowest $x$ bins and a negligible
      effect for the higher $x$ bins.
\item In the calculation of the acceptance, the fraction of the total
      cross section arising from diffractive scattering is increased
      by 60\%. The effect on $F_2$ is less than 5\%.
\item The efficiency
      for reconstructing a vertex is determined by MC simulation and is
      found to decrease from 85\% at $y=0.7$ to 40\% at $y=0.03$.
      For events at low $y$ the tracks in the CTD are at low angles
      as the current jet is produced
      in the forward direction, resulting in a reduced
      probability of reconstructing the event vertex.
      In the data, 70\% of the events have a tracking vertex
      compared to 73\% in the MC simulation.
      The vertex of the events without a tracking vertex
      is set to the nominal shifted vertex position.
      The reconstructed $Q^2$ values
      are larger than the true $Q^2$ values if the events
      without a tracking vertex originated from the satellite bunch.
      The effect on $F_2$
      of the satellite bunch was studied by MC simulation and
      it was found to be largest at high $x$ values where
      it is 8\% decreasing to about 1\% for the low $x$ points.

      As an additional check, the $Z$ coordinate of the vertex is fixed
      at the primary interaction point for all events in the
      data and in the MC simulation in order to check the
      sensitivity to the vertex distribution.
      The change in $F_2$ is largest for the high $Q^2$ values at large $x$
      where it
      is about 15\% of $F_2$. For the lowest $x$ and $Q^2$ bins the effect
      on $F_2$ is about 10\%, while for the other bins it is less than 5\%.
      The change in $F_2$ from this systematic check
      is not included when determining the systematic error.

\item The $\delta$ cut is lowered from 35 GeV to 32 GeV and raised to 38 GeV.
      The largest effect on $F_2$
      of about 3\% occurs at the low $x$ and low $Q^2$ bins,
      and it decreases with increasing $Q^2$ and $x$.
      Changing the $\delta$ cut checks the photoproduction background
      estimate, the simulation of the QED radiative effects and the
      simulation of the hadronic energy measurement.
\end{itemize}

The total systematic error on $F_2$ of this analysis is 4 to 14\% to be
compared with the statistical error of 4 to 9\%, which includes the
statistical error from the MC simulation (see Table~\ref{F2_table}).
In addition to the above errors, there is an overall normalisation
uncertainty of 3\% due to the uncertainty in the first level trigger
efficiency and to the uncertainty on the determination of the luminosity.
The given errors of the $F_2$ data do not include this normalisation
uncertainty.

\section{Analysis of the Radiative Events}

\label{sectisr}
The ISR analysis is based on data, corresponding to
an integrated luminosity of 2.5 ${\rm pb^{-1}}$, taken in 1994 with the
interaction point at $Z$~=~0. $F_2$ is measured
using events with QED initial state radiation. The emission of a hard
photon from the initial state positron results in a scattering at a
reduced centre-of-mass energy, and thus for a given scattering angle
events with lower $Q^2$ are accepted.

\subsection{Event Selection and Bin Selection}

The sample of DIS events with initial state radiation is selected
in a manner identical to the SVX analysis, with the
following exceptions:
\begin{itemize}
\item  An energy deposit in the LUMI photon calorimeter
       is demanded to ensure collinear radiation. The energy
       is required to be between 6 and 18 GeV in order to
       reduce background and ensure sufficient resolution of the
       reconstructed kinematic variables.
       Events having an energy greater than 3~GeV measured in the
       electron calorimeter of the luminosity detector are rejected
       in order to reduce the
       accidental overlap of DIS and photoproduction with those from
       the bremsstrahlung process $ep\rightarrow e\gamma p$.
\item  The corrected energy of the scattered positron detected in the RCAL
       is required
       to be greater than 8~GeV.  For positrons within the fiducial
       volume of the SRTD, the energy is corrected
       as in the SVX analysis.
       In areas not covered by the SRTD a position dependent description of
       the inactive material in front of the CAL is determined using the KP
       events and is used to correct the measured energy in the CAL.
       A description of the method can be found in~\cite{b:ZEUS_F2}.
       A correction for the positron
       identification efficiency in the MC simulation is also made as
       in the SVX analysis.
\item The fiducial cut for the positron position
       is the same as that described in the SVX
       analysis for the inner edge of the SRTD.
       If the positron position is outside the fiducial volume of the
       SRTD, the position is reconstructed using the calorimeter.
\item The $\delta$ cut is replaced by a cut on
       $\delta^\prime =  \delta + 2 \cdot E_\gamma$.
\end{itemize}

A sample of 10726 events is selected.
The resolutions of $x$ and $Q^2$ are similar to
those given in Sect.~\ref{s:resolution} and depend only
weakly on the photon energy $E_\gamma$.  The bins in $x$ and $Q^2$ are
chosen in a similar way to those used in the SVX analysis.

\subsection{Background Estimation}

The main source of background is the accidental overlap of
DIS or photoproduction events with bremsstrahlung events,
$e p \rightarrow e \gamma p $.
This background is estimated with events selected with the
above criteria, removing the requirement of a tagged photon
and the $\delta^\prime$ cut. To these events a photon is added with
an energy determined from a random sampling of the measured spectrum
of photon energies for
bremsstrahlung events. The resulting $\delta^\prime$ spectrum
is normalised to the $\delta^\prime$ distribution of the data above
70 GeV where no DIS events are possible, taking energy and momentum
conservation and the resolution of the detector into consideration,
to obtain an estimate of the background. It is subtracted from the data for
each $(x, Q^2)$ bin and is below 10\% of the number of events observed
in the bin except in the
lowest $x$ bins where it is up to 24\% in one bin.

The beam related background is determined and subtracted statistically
in the same manner as described in Sect.~\ref{s:kinematic_reconstruction}.
It is below 5\% for the three lowest $Q^2$ bins except for the bin
at the lowest $x$ where it is 7\%. It is negligible in higher $Q^2$ bins.
The background from photoproduction with initial state radiation
and cosmic rays is negligible. Event losses due to bremsstrahlung
overlap with a DIS event having initial state radiation are also negligible.

\subsection{Determination of $F_2$}

\begin{figure}[p]
\begin{center}
\mbox{\psfig{file=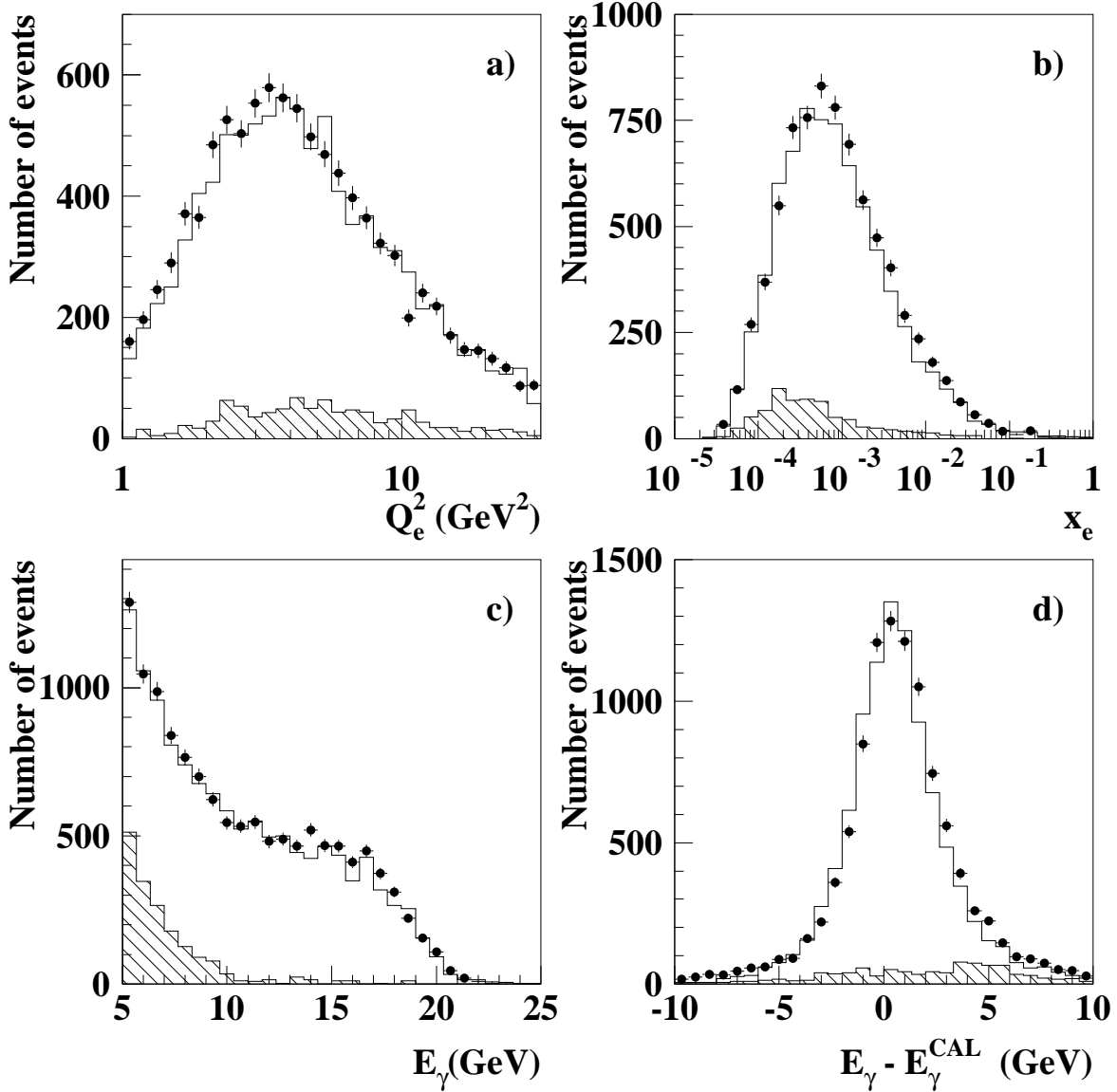,width=17cm}}
\end{center}
\caption[]
{\it
a) The reconstructed $Q^2_e$ distribution of the ISR sample.
b) The reconstructed $x_e$ distribution of the ISR sample.
c) The spectrum of the photon energy measured in the LUMI photon calorimeter
without the cut on the photon energy.
d) The difference between the photon energy measured in the LUMI photon
calorimeter and that determined from the CAL.
In the figures the data (dots),
background estimate (hatched histogram) and sum of the
background and DIS MC (solid histogram) are shown.
All events with a reconstructed $Q^2_e > 1$~${\rm GeV^2}$
which pass the selection criteria described in
Sect.~6.1 are shown.
The MC distributions have been reweighted using
the final $F_2$ parameterisation from the QCD NLO fit to the ZEUS data and
normalised to the luminosity of the data.}
\label{f:isr_figures}
\end{figure}

$F_2$ is determined as described in Sect.~\ref{s:determine_f2}.
The MC sample is first reweighted using the QCD~NLO fit to the $F_2$ from the
SVX analysis. $F_2$ is then determined using the reweighted MC
for the acceptance correction. A second QCD~NLO fit to the measured
ISR $F_2$ values (excluding the SVX $F_2$ values)
is performed and the final $F_2$ is determined.
The measured distributions of $Q^2$ and $x$ are shown in
Figs.~\ref{f:isr_figures}a and~\ref{f:isr_figures}b respectively.
The data are seen to be in adequate agreement with the MC simulation after the
background is added.

The photon energy spectrum of the selected events without the cut on
$E_\gamma$ is shown in
Fig.~\ref{f:isr_figures}c
and is well described by the expectation from the DIS MC simulation and
background estimates.  Figure~\ref{f:isr_figures}d shows the difference
between the energy of the radiated photon measured in the LUMI photon
calorimeter and $E_\gamma^{CAL}$,
the photon energy determined, using energy and momentum conservation,
from the measurements
of the energies and the angles of the scattered positron and hadrons in
the CAL.
The agreement between the MC simulation and the data
is reasonable. The distribution is centred at zero
indicating agreement in
the energy scales of the LUMI photon calorimeter and the CAL.

\subsection{Systematic Uncertainties}
The checks used to estimate the systematic uncertainty
are similar to those presented in Sect.~\ref{s:systematic_uncertainties}
except for the checks related to the displacement of the vertex.
To estimate the uncertainties due to the photon tagging
the following additional checks are made:
\begin{itemize}
\item The lower cut on the photon energy is raised from 6 to 7 GeV.
      The effect on $F_2$
      is below 2.5\% for the bins at low $y$ and around
      6\% for the bins at high $y$
      and is compatible with statistical fluctuations.
\item The resolution of the photon energy measurement in the MC simulation
      is degraded from $26.5\%\cdot\sqrt{E \, ({\rm GeV})}$
      to $28.5\%\cdot\sqrt{E \, ({\rm GeV})}$. The effect on $F_2$
      is around 1.5\%.
\item The energy calibration of the LUMI photon calorimeter in
      the MC simulation is changed linearly, within the estimated
      uncertainties, by
      $0.4 \%$ at the positron beam energy to $3 \%$ at 5~GeV. The effect
      on $F_2$ is around 7\%.
\item The acceptance of the LUMI photon calorimeter for ISR events obtained
      from the MC simulation is about
      30\%. An independent determination of the acceptance using
      bremsstrahlung data leads to an effect of about 4\% on $F_2$.

      The beam divergence is determined with the LUMI photon calorimeter
      using the bremsstrahlung events.
      If the beam divergence is increased
      by 15\% in the acceptance calculation,
      the effect on $F_2$ is, on average, 1.7\%.
\item The event selection is repeated without the requirement that the
      LUMI electron energy be less than 3 GeV.  The effect
      is below 3\% for the bins at low $y$ and is up to 10\% in two
      bins at high $y$.
\end{itemize}

The total systematic uncertainty of this analysis is 14 to 27\% to be compared
to the statistical error of 7 to 14\%.
The effect of additional QED radiative corrections not included in the
HERACLES MC programme is small relative to the measurement errors
in the kinematic region considered here \cite{b:SP}.

\section{Results and Discussion}
\begin{figure}[p]
\begin{center}
\mbox{\psfig{file=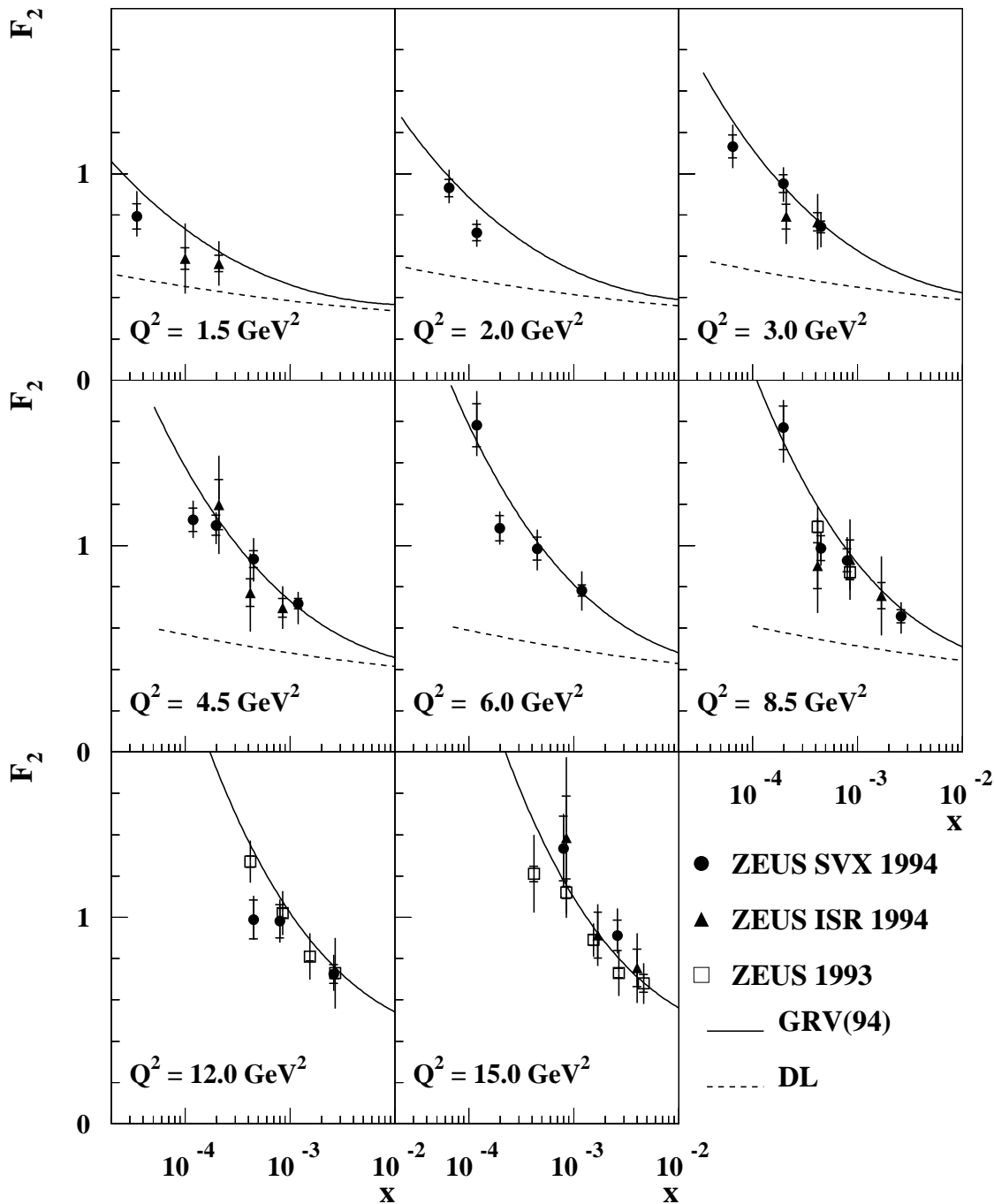,width=16cm}}
\end{center}
\caption[kkk]
{\it
The measured $F_2$ from the SVX analysis (solid dots), the
ISR analysis (solid triangles) and the 1993 results (open squares)
compared with the expectations from GRV(94) (solid line) and
Donnachie and Landshoff (DL) (dashed line).
Overall normalisation uncertainties of 3\% for the
1994 results and 3.5\% for the 1993 points are not shown.
The inner error bars represent the statistical errors while the outer
error bars represent the systematic errors added in quadrature to
the statistical errors.
}
\label{fig_f2_curve}
\end{figure}

The measured $F_2$ values from the SVX data
(ZEUS SVX 1994) are listed in Table~\ref{F2_table}, and
those from the ISR analysis (ZEUS ISR 1994)
are listed in Table~\ref{F2isr_table}.
The data are presented in Fig.~\ref{fig_f2_curve} together
with previous ZEUS measurements (ZEUS 1993).
The three data sets are in good agreement.
The present analyses significantly increase the measured kinematic region.
Using our data alone, the rise of $F_2$ as $x$ decreases is observed down to
$Q^2$~=~3~${\rm GeV^2}$. Including the fixed target data
(the $F_2$ parameterisations shown as curves in Fig.~\ref{fig_f2_curve}
provide a good description of the fixed target data at higher $x$)
the rise is
seen to persist down to $Q^2$~=~1.5~${\rm GeV^2}$.

In Fig.~\ref{fig_f2_curve} the measured $F_2$ values
are compared to the prediction of GRV(94)~\cite{b:GRV94} which is
based on perturbative QCD using the DGLAP evolution equations.
The GRV(94) parton distributions have a very low starting scale, $Q_{\rm o}^2$,
for the DGLAP evolution equation of $0.34~{\rm GeV}^2$,
where the gluon and sea distributions are assumed to have
valence-like spectra.
The steep rise in $F_2$ at low $x$ is generated dynamically by the
evolution in $Q^2$. The GRV(94) predictions for $F_2$ are in agreement
with the data showing that perturbative QCD can describe the data
down to $Q^2$ values of 1.5 ${\rm GeV^2}$ at the low $x$ values of
this measurement.
The predictions of Donnachie and Landshoff~\cite{b:DL}
based on Regge phenomenology are also shown and are seen to be
ruled out for $Q^2\ge2$~${\rm GeV^2}$ and disfavoured for
$Q^2 = 1.5$~${\rm GeV^2}$.

\begin{figure}[p]
\begin{center}
\mbox{\psfig{file=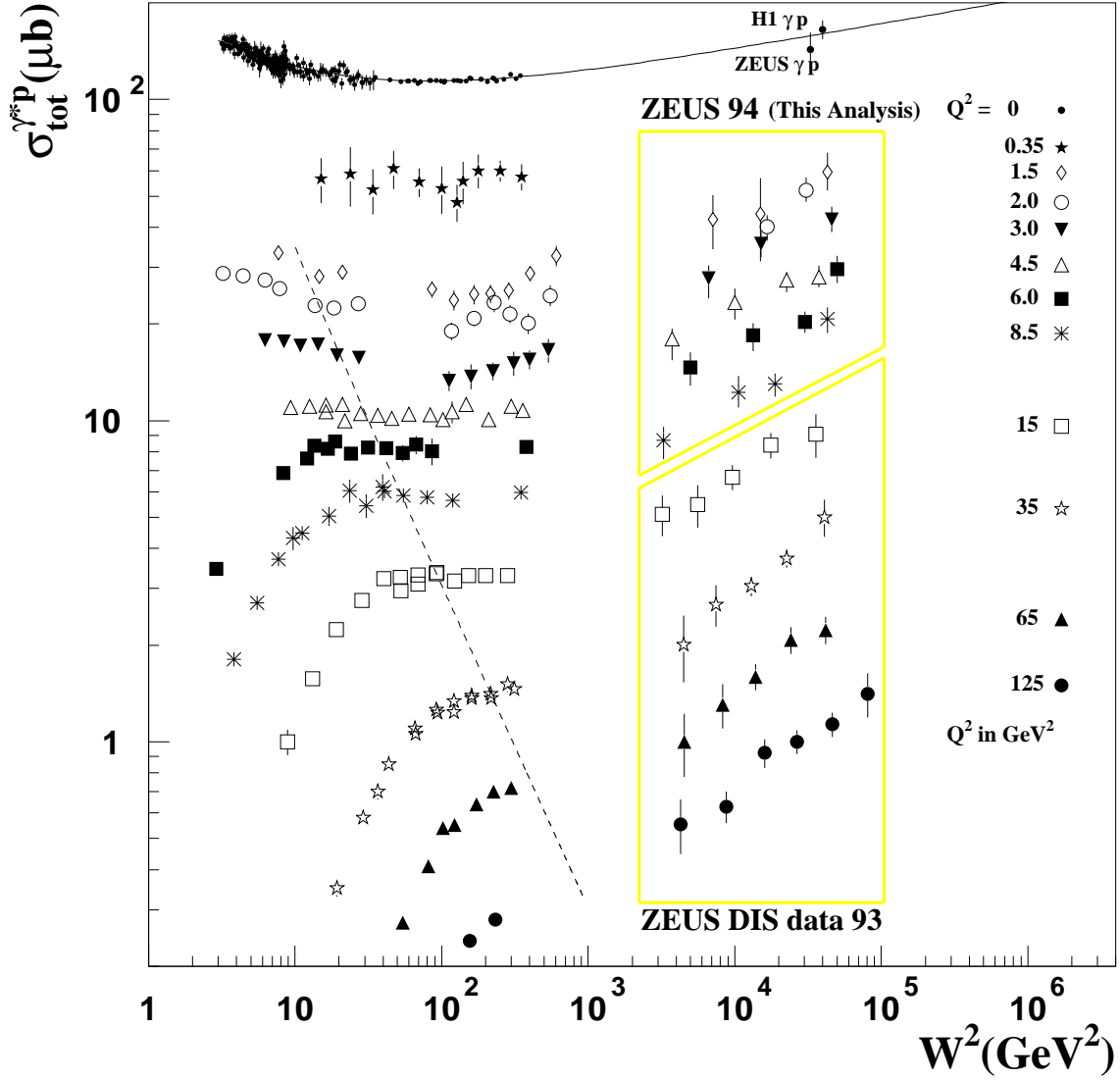,width=15cm}}
\end{center}
\caption[]
{\it
The total virtual photon-proton cross section versus
$W^2$ for different $Q^2$ values. The cross section values obtained from
the $F_2$ values described in this paper and the $F_2$ values from
the 1993 data are shown in addition to data from previous low energy
experiments \cite{b:fixed}.
The region to the right of the dashed line correspond to
$x < 1/(2 m_p R_p)$.
Also shown is the $W^2$ behaviour of the measured cross section for
real photoproduction together with the prediction
of Donnachie and Landshoff~\cite{b:DL_photo} (solid line).
}
\label{f:sigtot}
\end{figure}

The DIS cross section can be expressed as the product of the flux of virtual
photons and the total cross section $\sigma_{tot}^{\gamma^* p}$ for
the scattering of virtual photons on protons~\cite{b:HAND}.
$\sigma_{tot}^{\gamma^* p}$ is defined in terms of the cross section
for the absorption of transverse and longitudinal photons, $\sigma_T$ and
$\sigma_L$ respectively, by
\begin{equation}
   \sigma_{tot}^{\gamma^* p} \equiv \sigma_T(x,Q^2) + \sigma_L(x,Q^2).
\end{equation}
The expression for $F_2$ in terms of $\sigma_T$ and $\sigma_L$ is
\begin{equation}
 F_2(x,Q^2)=\frac{Q^2(1-x)}{4\pi^2\alpha}
            \frac{Q^2}{Q^2+4m_p^2x^2} [\sigma_T(x,Q^2)+\sigma_L(x,Q^2)],
\end{equation}
where $m_p$ is the mass of the proton.
The separation into the photon flux and cross section can be interpreted
in a way similar to the interaction  of real particles provided that the
lifetime of the virtual photon is large compared to the interaction time,
or $x \ll 1/(2 m_p R_p)$ where
$R_p \approx 4$~${\rm GeV^{-1}}$ is the proton radius~\cite{b:p_radius}.
At small $x$ the expression can be
written in terms of the total virtual photon-proton centre-of-mass energy $W$
(where $W^2 = m_p^2 + Q^2(1/x-1)$) to give
\begin{equation}
\label{e:sigtot}
\sigma_{tot}^{\gamma^*p}(W^2,Q^2) \approx \frac{4 \pi^2 \alpha}{Q^2}F_2(x,Q^2).
\end{equation}
The measured $F_2$ data are converted to the total virtual photon-proton
cross section and shown in Fig.~\ref{f:sigtot} along with low
energy data and  real photoproduction cross
section measurements. The DIS data are seen to rise steeply
as a function of $W^2$ between the fixed target and the HERA energy range,
even at $Q^2$ values as low as $Q^2 = 2.0$ ${\rm GeV^2}$ in contrast
to the cross section for real photons which exhibits
only a slow rise from the fixed target data to the HERA
data~\cite{b:sigtot_photoprod,b:H1_photoprod}.

In summary, the proton structure function $F_2(x,Q^2)$ has been
measured in DIS in a new kinematic range down
to $Q^2 = 1.5$~${\rm GeV^2}$ and $x = 3.5 \times 10^{-5}$.
The GRV(94) predictions, which are based on perturbative QCD,
are found to be consistent with the data.
The Regge predictions of Donnachie and Landshoff based on the soft pomeron
are ruled out for $Q^2 \ge$~2~${\rm GeV^2}$ and disfavoured for
$Q^2 = 1.5$~${\rm GeV^2}$.
For the centre-of-mass energies between the fixed target regime and
the HERA energy range,
the total virtual photon-proton cross section $\sigma_{tot}^{\gamma^* p}$
is found to rise steeply with
the centre-of-mass energy for $Q^2$ as low as 2.0~${\rm GeV^2}$.

\section{Acknowledgements}
The
strong support and encouragement of the DESY Directorate has
been invaluable.
The experiment was made possible by the inventiveness and the diligent
efforts of the HERA machine group.  The design, construction and
installation of the ZEUS detector have been made possible by the
ingenuity and dedicated efforts of many people from inside DESY and
from the home institutes who are not listed as authors. Their
contributions are acknowledged with great appreciation.

\begin{table}[p]
\centering
{\footnotesize
\begin{tabular}{|r|r|r|r|r|r|r|r|}
\hline
\mc{$Q^2$} & \mc{$Q^2$ range} & \mc{$x$} & \mc{$x$ range} & \mc{No.} & \mc{No.
BG}  & \mc{$R$} & \mc{Measured} \\
\mc{${\rm GeV^2}$} & \mc{${\rm GeV^2}$} &     &  & \mc{events} & \mc{events} &
    &
   \mc{$F_2 \pm \rm{stat} \pm \rm{sys}$} \\
\hline \hline
1.5  & 1.3 - 1.9 & $3.5 \cdot 10^{-5}$ & $2.8 - 5.2  \cdot 10^{-5}$  & 292  &
14.6 & 0.71 & 0.79 $\pm$ 0.06 $\pm^{0.11}_{0.08}$ \\ \hline
2.0  & 1.9 - 2.7 & $6.5 \cdot 10^{-5}$ & $4.0 - 9.0  \cdot 10^{-5}$  & 747  &
20.0 & 0.59 & 0.93 $\pm$ 0.04 $\pm^{0.08}_{0.06}$ \\
     &           & $1.2 \cdot 10^{-4}$ & $0.9 - 2.0  \cdot 10^{-4}$  & 456  &
2.7 & 0.53 & 0.71 $\pm$ 0.04 $\pm^{0.05}_{0.06}$ \\ \hline
3.0  & 2.7 - 3.6 & $6.5 \cdot 10^{-5}$ & $0.58 - 1.2 \cdot 10^{-4}$  & 600  &
17.3 & 0.57 & 1.13 $\pm$ 0.06 $\pm^{0.09}_{0.09}$ \\
     &           & $2.0 \cdot 10^{-4}$ & $1.2 - 2.3  \cdot 10^{-4}$  & 706  &
1.3 & 0.49 & 0.95 $\pm$ 0.04 $\pm^{0.07}_{0.08}$ \\
     &           & $4.5 \cdot 10^{-4}$ & $0.23 - 1.0 \cdot 10^{-3}$  &1067  &
 0 & 0.44 & 0.74 $\pm$ 0.03 $\pm^{0.06}_{0.10}$ \\ \hline
4.5  & 3.6 - 5   & $1.2 \cdot 10^{-4}$ & $0.8 - 1.57 \cdot 10^{-4}$  & 503  &
12.0 & 0.52 & 1.12 $\pm$ 0.06 $\pm^{0.08}_{0.07}$ \\
     &           & $2.0 \cdot 10^{-4}$ & $1.57 - 3.0 \cdot 10^{-4}$  & 643  &
1.3 & 0.50 & 1.10 $\pm$ 0.05 $\pm^{0.04}_{0.08}$ \\
     &           & $4.5 \cdot 10^{-4}$ & $3.0 - 6.0  \cdot 10^{-4}$  & 705  &
 0 & 0.45 & 0.93 $\pm$ 0.04 $\pm^{0.09}_{0.10}$ \\
     &           & $1.2 \cdot 10^{-3}$ & $0.6 - 4.0  \cdot 10^{-3}$  &1148  &
 0 & 0.39 & 0.72 $\pm$ 0.03 $\pm^{0.05}_{0.10}$ \\ \hline
6.0  & 5 - 7     & $1.2 \cdot 10^{-4}$ & $1.1 - 1.8  \cdot 10^{-4}$  & 289  &
8.0 & 0.53 & 1.58 $\pm$ 0.10 $\pm^{0.13}_{0.11}$ \\
     &           & $2.0 \cdot 10^{-4}$ & $1.8 - 3.2  \cdot 10^{-4}$  & 389  &
1.3 & 0.50 & 1.08 $\pm$ 0.06 $\pm^{0.05}_{0.05}$ \\
     &           & $4.5 \cdot 10^{-4}$ & $3.2 - 5.6  \cdot 10^{-4}$  & 410  &
 0 & 0.46 & 0.98 $\pm$ 0.06 $\pm^{0.07}_{0.09}$ \\
     &           & $1.2 \cdot 10^{-3}$ & $0.56 - 3.0 \cdot 10^{-3}$  &1016  &
 0 & 0.40 & 0.78 $\pm$ 0.03 $\pm^{0.09}_{0.09}$ \\ \hline
8.5  & 7 - 10    & $2.0 \cdot 10^{-4}$ & $1.5 - 3.0  \cdot 10^{-4}$  & 268  &
4.0 & 0.51 & 1.57 $\pm$ 0.11 $\pm^{0.08}_{0.13}$ \\
     &           & $4.5 \cdot 10^{-4}$ & $3.0 - 6.0  \cdot 10^{-4}$  & 314  &
1.3 & 0.47 & 0.99 $\pm$ 0.06 $\pm^{0.05}_{0.06}$ \\
     &           & $8.0 \cdot 10^{-4}$ & $0.6 - 1.2  \cdot 10^{-3}$  & 333  &
 0 & 0.44 & 0.93 $\pm$ 0.06 $\pm^{0.10}_{0.08}$ \\
     &           & $2.6 \cdot 10^{-3}$ & $1.2 - 7.0  \cdot 10^{-3}$  & 542  &
 0 & 0.36 & 0.66 $\pm$ 0.03 $\pm^{0.06}_{0.08}$ \\ \hline
12.0 & 10 - 14   & $4.5 \cdot 10^{-4}$ & $2.5 - 6.0  \cdot 10^{-4}$  & 124  &
2.7 & 0.48 & 0.99 $\pm$ 0.09 $\pm^{0.07}_{0.03}$ \\
     &           & $8.0 \cdot 10^{-4}$ & $0.6 - 1.2  \cdot 10^{-3}$  & 167  &
 0 & 0.45 & 0.98 $\pm$ 0.08 $\pm^{0.07}_{0.07}$ \\
     &           & $2.6 \cdot 10^{-3}$ & $1.2 - 5.0  \cdot 10^{-3}$  & 291  &
 0 & 0.37 & 0.73 $\pm$ 0.05 $\pm^{0.08}_{0.07}$ \\ \hline
15.0 & 14 - 20   & $8.0 \cdot 10^{-4}$ & $0.6 - 1.5  \cdot 10^{-3}$  &  81  &
 0 & 0.46 & 1.33 $\pm$ 0.16 $\pm^{0.06}_{0.10}$ \\
     &           & $2.6 \cdot 10^{-3}$ & $0.15 - 1.2 \cdot 10^{-2}$  & 171  &
 0 & 0.38 & 0.91 $\pm$ 0.07 $\pm^{0.11}_{0.04}$ \\
\hline
\end{tabular}
}
\caption[F2_table]
{\it The measured $F_2(x,Q^2)$ from the SVX analysis. The
bin boundaries and values of $x$ and $Q^2$ at which
$F_2$ is determined are listed.
The numbers of events
before background subtraction as well as the
estimated photoproduction and beam-related background
(in the column labeled ``No. BG events'')
for each bin are given.
The values of $R$, which are used to determine $F_2$ from the differential
cross sections, are shown (see text). An overall normalisation error of 3\% is
not
included.
}
\label{F2_table}
\end{table}

\begin{table}[p]
\centering
{\footnotesize
\begin{tabular}{|r|r|r|r|r|r|r|r|}
\hline
\mc{$Q^2$} & \mc{$Q^2$ range} & \mc{$x$} & \mc{$x$ range} & \mc{No.} & \mc{No.
BG}  & \mc{$R$} & \mc{Measured} \\
\mc{${\rm GeV^2}$} & \mc{${\rm GeV^2}$} &     &  & \mc{events} & \mc{events} &
    &
   \mc{$F_2 \pm \rm{stat} \pm \rm{sys}$} \\
\hline \hline
1.5 & 1.3 - 2.2 & $1.0 \cdot 10^{-4}$ & $0.65 - 1.5 \cdot 10^{-4}$ & 273 & 34.4
& 0.72 & 0.59 $\pm$ 0.05 $\pm$ 0.16 \\
    &           & $2.1 \cdot 10^{-4}$ & $1.5 - 4.5 \cdot 10^{-4}$ & 474 & 16.5
&  0.61 & 0.57 $\pm$ 0.04 $\pm$ 0.10 \\ \hline
3.0 & 2.2 - 3.8 & $2.1 \cdot 10^{-4}$ & $1.5 - 3.0 \cdot 10^{-4}$ & 429 & 68.5
&  0.57 & 0.79 $\pm$ 0.06 $\pm$ 0.12 \\
   &            & $4.2 \cdot 10^{-4}$ & $3.0 - 9.0 \cdot 10^{-4}$ & 695 & 49.5
&  0.52 & 0.77 $\pm$ 0.05 $\pm$ 0.13 \\ \hline
4.5 & 3.8 - 6.5 & $2.1 \cdot 10^{-4}$ & $1.5 - 3.0 \cdot 10^{-4}$ & 265 & 55.7
&  0.56 & 1.20 $\pm$ 0.12 $\pm$ 0.21 \\
   &            & $4.2 \cdot 10^{-4}$ & $3.0 - 6.0 \cdot 10^{-4}$ & 340 & 59.8
&  0.51 & 0.77 $\pm$ 0.07 $\pm$ 0.18 \\
   &            & $8.5 \cdot 10^{-4}$ & $0.6 - 1.8 \cdot 10^{-3}$ & 570 & 42.5
&  0.47 & 0.70 $\pm$ 0.05 $\pm$ 0.09 \\ \hline
8.5 & 6.5 - 11.5& $4.2 \cdot 10^{-4}$ & $3.0 - 6.0 \cdot 10^{-4}$ & 189 & 45.6
&  0.52 & 0.90 $\pm$ 0.11 $\pm$ 0.20 \\
   &            & $8.5 \cdot 10^{-4}$ & $0.6 - 1.2 \cdot 10^{-3}$ & 246 & 31.2
&  0.48 & 0.93 $\pm$ 0.10 $\pm$ 0.17 \\
   &            & $1.7 \cdot 10^{-3}$ & $1.2 - 3.6 \cdot 10^{-3}$ & 341 & 15.0
&  0.43 & 0.76 $\pm$ 0.06 $\pm$ 0.18 \\ \hline
15.0 & 11.5 - 20& $8.5 \cdot 10^{-4}$ & $0.6 - 1.2 \cdot 10^{-3}$ & 143 & 22.5
&  0.49 & 1.39 $\pm$ 0.20 $\pm$ 0.33 \\
   &            & $1.7 \cdot 10^{-3}$ & $1.2 - 2.4 \cdot 10^{-3}$ & 159 & 8.7 &
 0.45 & 0.91 $\pm$ 0.11 $\pm$ 0.10 \\
   &            & $4.0 \cdot 10^{-3}$ & $2.4 - 7.2 \cdot 10^{-3}$ & 178 & 9.8 &
 0.38 & 0.76 $\pm$ 0.09 $\pm$ 0.14 \\ \hline
\end{tabular}
}
\caption[F2isr_table]
{\it The measured $F_2(x,Q^2)$ from the ISR analysis.
The bin boundaries and values, of $x$ and $Q^2$ at which
$F_2$ is determined are listed.
The numbers of events before background subtraction as well as the
estimated accidental event overlap and beam-related background
(in the column labeled ``No. BG events'')
for each bin are given.
The value of $R$ obtained from the NLO QCD fit
is tabulated.
An overall normalisation error of 3\% is not included.
}
\label{F2isr_table}
\end{table}

\end{document}